\documentclass[%
reprint,
superscriptaddress,
amsmath,amssymb,
aps,
prb,
]{revtex4-2}

\usepackage{graphicx}% Include figure files
\usepackage{dcolumn}% Align table columns on decimal point
\usepackage{bm}% bold math
\usepackage{mathtools}
\usepackage{stackrel}

\usepackage{siunitx}
\newcommand{\um}{\SI{}{\micro\metre}}

\begin{document}

%\preprint{APS/123-QED}

\title{Free electron nonlinearities in heavily doped semiconductors plasmonics}

\author{Federico De Luca}
\affiliation{Istituto Italiano di Tecnologia, Center for Biomolecular Nanotechnologies, Via Barsanti 14, 73010 Arnesano, Italy}
\affiliation{Dipartimento di Matematica e Fisica "E. De Giorgi", Universit\`a del Salento, via Arnesano, 73100 Lecce, Italy}
\author{Michele Ortolani}
\affiliation{Department of Physics, Sapienza University of Rome, Piazzale Aldo Moro 5, 00185 Rome, Italy}
\affiliation{Istituto Italiano di Tecnologia, Center for Life NanoSciences, Viale Regina Elena 291, 00161 Rome, Italy}
\author{Cristian Cirac\`i}%
 \email{cristian.ciraci@iit.it}
\affiliation{Istituto Italiano di Tecnologia, Center for Biomolecular Nanotechnologies, Via Barsanti 14, 73010 Arnesano, Italy}

\date{\today}% It is always \today, today,

\begin{abstract}
Heavily doped semiconductors have emerged as tunable low-loss plasmonic materials at mid-infrared frequencies.
In this  article we investigate nonlinear optical phenomena associated with high concentration of free electrons.  We use a hydrodynamic description to study  free electron  dynamics in heavily doped semiconductors up to third-order  terms,  which are usually negligible for noble metals.
We find that cascaded third-harmonic generation due to second-harmonic  signals  can  be  as  strong  as  direct  third-harmonic generation contributions even when the second-harmonic generation efficiency is zero.
Moreover, we show that when coupled with plasmonic enhancement free electron nonlinearities could be  up  to  two  orders  of  magnitude  larger  than  conventional  semiconductor nonlinearities.
Our study might open a new route for nonlinear optical integrated devices at mid-infrared frequencies.
\end{abstract}

\maketitle

%\tableofcontents

\section{\label{sec:intro}Introduction}
Plasmonic nanoantennas emerged as a disruptive technology for concentrating and controlling light at the nanoscale thanks to their unique capabilities of field localization and enhancement \cite{Baumberg:2019iw,Ojambati:2019kv,Huang:2019jx,Chen:2018ky,FernandezDominguez:2018cw,Koenderink:2017cg}.
Nonetheless, their widespread application has struggled to concretize because of several shortcomings of the main constituent of plasmonic materials: noble metals.
The use of noble metals for practical devices has been limited due to, above all, high losses, high reflectivity and very poor compatibility with standard microelectronics manufacturing processes.
In this context, heavily doped semiconductors (i.e. with charge densities $n_0\sim 10^{19}$--$10^{20}$~cm$^{-3}$)  have been introduced as low-loss and tunable alternative materials for plasmonics in the near-infrared (NIR), i.e. $0.8<\lambda<2$~\um, and in the mid-infrared (MIR), i.e. $2<\lambda<20$~\um ~\cite{Boltasseva:2011ep, Naik:2013am}, being $\lambda$ the free-space wavelength.
The MIR overlaps with the so-called molecular fingerprint region, and is particularly important for sensing applications in many fields, from healthcare to security \cite{Taliercio:2019}.
Moreover, interest in MIR devices is growing in the context of free-space communications, since the 3--5~\um~and the 8--13~\um~regions are less affected by atmospheric turbulence, clouds and fog than the NIR region \cite{Su:2018fa}.
Because semiconductors offer the possibility of tuning their optical response through doping or electrical and optical excitation, and some of them are also highly compatible with silicon CMOS technology, their use for plasmonics may lead to breakthrough technologies \cite{Taliercio:2019}.  

Plasmonic systems owe their properties to surface plasmon modes -- the resonant collective oscillations of free electrons (FEs)-- that appear in materials with a high carrier concentration (i.e. metals and heavily doped semiconductors) as a consequence of the interaction with an external light excitation.
Naturally, the local field enhancement provided by these modes especially affects optical processes whose dependence on the field intensity is nonlinear.
Indeed, this is extremely important for nonlinear optical effects since they are in general very weak and normally require high laser intensities and long propagation distances in macroscopic nonlinear crystals to reach operational efficiencies.
Although, recently, high-index dielectric resonators have been used to partially overcome these limitations \cite{Koshelev:2020jm,Gili:2016hl,Carletti:2018gc}, 
plasmonic nanoantennas made of noble metals have been commonly used as local field amplifier in hybrid systems to boost optical nonlinearity from dielectric material placed in their vicinity \cite{QixinShen:2020cw,Sarma:2019im,Zeng:2018bp,Echarri:2018fi,Shen:2018dy,Guddala:2016ik,Shibanuma:2017kv,Lee:2014ewa,Wang:2017gw,Noor:2020kq,Hentschel:2016er,Barakat:2012bi,Deng:2020gq,Nielsen:2017du}.
The nonlinear response may also originate directly from the plasmonic material itself \cite{Harutyunyan:2012jj,Ko:2011cha,Czaplicki:2018dca} and, in particular, from the dynamics of non-equilibrium FEs  \cite{Krasavin:2017ki,Chizmeshya:1988ft, Crouseilles:2008hv, Scalora:2010kd, Pavlyukh:2012ei,Ciraci:2012vk}.
For instance, in noble metals, FE plasmonic nonlinearities have been shown to strongly contribute to second-order processes \cite{Scalora:2010kd,Capretti:2013bq,Wells:2018dc} in the visible and NIR, and experimental measurements in gold nanoparticle arrays have demonstrated second-harmonic generation (SHG) efficiencies comparable to those of nonlinear crystals when normalized to the active volumes \cite{Klein:2006tj,Czaplicki:2013}.

Although FE optical nonlinearities have mostly been observed in metals, analogous effects may also occurs in heavily doped semiconductors. 
The dynamic properties of doped semiconductors undergo in fact an interesting transition from the size-quantization regime to the classical regime of plasmon oscillations.
In the former regime, semiconductor quantum-wells have enabled the strongest nonlinear susceptibilities \cite{Lee:2014ew}.
The resonant nature of electron transitions in quantum-wells however severely hinders their application in ultra-fast devices, since it inevitably causes a slower response and increased decoherence.
The plasmonic regime for semiconductors, on the other hand, unlocks a new realm of possible ultra-fast nonlinear effects \cite{Fischer:2018bs, Wagner:2014}, and offers several advantages with respect to nonlinear plasmonics in noble metals. 

To understand the advantage of using heavily doped semiconductors, it should be noticed that within the hydrodynamic description, third order nonlinear effects are inversely proportional to the square of the equilibrium free-carriers density, i.e. $\mathbf{P}^{(3)}_{\rm NL} \propto \frac{1}{n_0^2}$ (as we will show later), where $\mathbf{P}^{(3)}_{\rm NL}$ is the third-order polarization vector.
In gold, for examples, the high concentration of free-carriers ($n_0\sim10^{22}$~cm$^{-3}$), leads to very weak contributions to the third-order nonlinear polarization if compared to contributions due to the crystal lattice nonlinear susceptibility $\chi^{(3)}$.
On the other hand, doped semiconductors with a plasma frequency in the MIR have a much lower charge density ($n_0\sim10^{19}$~cm$^{-3}$) than noble metals, such that nonlinear terms may grow as much as 6 order of magnitude, overcoming by far the  lattice nonlinearities.
Moreover, semiconductors have an effective electron mass, $m$, that can be one order of magnitude smaller than that of noble metals.
These characteristics cause also an increase of the nonlinear active volumes, whose measure (normalized to the operation wavelength) can be linked \cite{Maack:2017fn,Dias:2018bp} to the Fermi velocity $v_F \propto \frac{n_0^{1/3}}{m}$.

The large variety of semiconductors with different effective electron mass and doping levels give access to a wide range of possibilities to optimize and increase intrinsic nonlinear effects.
Furthermore, these quantities, unlike in metals, can also be used to control their optical linear properties.
Indeed, assuming a Drude model, a doped semiconductor is characterized by a dielectric function $\varepsilon (\omega) = \varepsilon_{\infty} - \frac{\omega_P^2}{\omega^2+i\gamma\omega}$, where $\varepsilon_{\infty}$ is the background permittivity, $\gamma$ is the damping rate and $\omega_{\rm p} = \sqrt{\frac{e^2n_0}{\varepsilon_0m}}$ is the plasma frequency of the material, being $e$ and $\varepsilon_0$ the elementary charge (absolute value) and the dielectric constant of vacuum (note that hole-doping is unsuitable for plasmonics due to inter-valence-band transitions taking place in the MIR).
By choosing the semiconductor, i.e. $m$ and the doping level, $n_0$, is it then possible to tune the dielectric function at will \cite{frigerio:2016}.
For example, one could generate an epsilon-near-zero (ENZ) medium, to explore unconventional nonlinear optical effects \cite{Reshef:2019,Deng:2020gq}.
For all these reasons, the study of nonlinear FE dynamics in heavily doped semiconductors seems to represent a very promising test field toward the understanding of the fundamental mechanisms of nonlinearity in condensed matter and therefore in the direction of designing innovative optical technologies.

In this article,  we focus on the process of third-harmonic generation (THG) from heavily doped semiconductors.
We use a hydrodynamic description to study the free carriers dynamics up to third order terms, usually neglected for noble metals, and consider both cascaded and direct contributions.
We first investigate THG efficiency in unpatterned doped semiconductor slabs.
Interestingly, we find that cascaded THG due to SHG signals can be as strong as direct THG contributions even when the SHG efficiency is zero, i.e., there is no SHG in the far-field.
We also suggest a possible experiment to demonstrate such effect. 
We then analyze the conditions that allow for the maximum THG efficiency and show that FE nonlinearities can be up to two orders of magnitude larger than intrinsic lattice nonlinearities.
We finally introduce a grating pattern on the semiconductor slab surface that allows to strongly couple with normally incident waves and excite local plasmons to further enhance the conversion efficiency.

%%%%%%%%%%%%%%%%%%%%%%%%%%%%%%%%%%%%%%%%%%%%%%%%%%%%%%%%%%%%%%%%%%%%%%%%%%%%%%%%%%%%%%%%%%%%%%
%%%%%%%%%%%%%%%%%%%%%%%%%%%%%%%%%%%%%%%%%%%%%%%%%%%%%%%%%%%%%%%%%%%%%%%%%%%%%%%%%%%%%%%%%%%%%%
\section{\label{sec:model}Hydrodynamic model for THG}
To model the many-body dynamics of interest, a theoretical framework that describes not only the charge density distribution and the nonlinear effects, but also nonlocal corrections, is needed.
Indeed, because of the fermionic nature of the carriers, FE nonlinearities are intrinsically nonlocal, in the sense that the induced currents depend not only on the value of the electric field at a given point but also, through their spatial derivatives, on the value of the fields in the surrounding area. 
In addition, the formalism should give the possibility to compare FE nonlinearities to those arising from the crystal lattice, in order to solve a possible ambiguity in the origin of the nonlinear response.
An appropriate representation of nonlinear and nonlocal FE dynamics in heavily doped semiconductors may be done taking into consideration the quasi-classical formalism of the hydrodynamic model, in the limit of Thomas-Fermi approximation \cite{Scalora:2020dq,Maack:2017fn,Rodriguez:2020s}.
The hydrodynamic theory has been successfully used to describe the FE dynamics in noble metal systems and is then extremely relevant to describe the behavior of heavily doped semiconductors.
Furthermore, lattice nonlinearities can be easily included into the model by taking into consideration an intrinsic nonlinear susceptibility, as shown in the following.

Within the hydrodynamic formalism \cite{Yan:2015ff, deCeglia:2018ip, Raza:2015ef}, the many-body dynamics of a FE fluid can be described taking into account two macroscopic variables, the charge density, $n(\mathbf{r},t)$, and the current density, $\mathbf{J}(\mathbf{r},t) = -en\mathbf{v}$, where $\mathbf{v}(\mathbf{r},t)$ is the electron velocity field.
Under the influence of external electric and magnetic fields, $\mathbf{E}(\mathbf{r},t)$ and $\mathbf{H}(\mathbf{r},t)$, the system can be described by the following equation:

\begin{eqnarray}
\label{eqn:HM}
m\left(\frac{\partial}{\partial t} -\frac{\mathbf{J}}{en}\cdot \nabla + \gamma\right)\frac{\mathbf{J}}{en} =&& +e\mathbf{E} -\frac{\mathbf{J}}{n} \times \mu_0\mathbf{H}+ \nonumber \\ 
&&+\nabla \frac{\delta T^{\rm TF}[n]}{\delta n},
\end{eqnarray}
where $\mu_0$ is the magnetic permeability of vacuum. The last term takes into account the nonlocal effects due to the quantum pressure. Here, $T^{\rm TF}[n]$ is the kinetic energy functional in the Thomas-Fermi approximation, whose derivative with respect to $n$ is $\frac{\delta T^{\rm TF}[n]}{\delta n}= c_{\rm TF}\frac{5}{3}n^{\frac{2}{3}}$, with $c_{\rm TF} = \frac{\hbar^2}{m}\frac{3}{10}(3\pi^2)^{2/3}$ \cite{Ciraci:2016il}.

Following a perturbative approach, it is possible to write $n({\bf r},t)=n_0+n_{\rm ind}({\bf r},t)$, where $n_{\rm ind}=\frac{1}{e}\nabla \cdot{\bf P}$ is the induced charge density and $n_{\rm ind} \ll n_0 $. Here ${\bf P}({\bf r},t)$ is the polarization vector, which is related to the current density through its time derivative, i.e. $\dot{\bf P}={\bf J}$.  Eq.~\eqref{eqn:HM} then becomes:
\begin{eqnarray}
\label{eqn:HM_P}
\ddot{\mathbf{P}}+\gamma\dot{\mathbf{P}} = \frac{n_0e^2}{m}{\mathbf{E}} +\beta^2\nabla(\nabla\cdot\mathbf{P})+\mathbf{S}_{\rm NL}^{(2)}+\mathbf{S}_{\rm NL}^{(3)}, 
\end{eqnarray}
where time derivatives are now expressed in dot notation.
This equation is essentially constituted by the FE gas linear model, i.e. the Drude model plus a nonlocal linear term, where $\beta^2=\frac{10}{9}\frac{c_{\rm TF}}{m}n_0^{2/3}$, and the second- and third-order nonlinear sources, $\mathbf{S}_{\rm NL}^{(2)}$ and $\mathbf{S}_{\rm NL}^{(3)}$, respectively, whose expressions are:
\begin{subequations}
\label{eqn:NL_sources}
\begin{eqnarray}
\label{eqn:2 order}
\mathbf{S}_{\rm NL}^{(2)}=&& \frac{e}{m}{\mathbf{E}}\nabla  \cdot {\mathbf{P}} - \frac{e\mu_0}{m}\dot{\mathbf{P}} \times {\mathbf{H}}+\frac{1}{{e{n_0}}}( {\dot{\mathbf{P}}\nabla  \cdot \dot{\mathbf{P}} + \dot{\mathbf{P}} \cdot \nabla \dot{\mathbf{P}}})\nonumber\\
&&+\frac{1}{3}\frac{\beta^2}{en_0}\nabla (\nabla  \cdot {\mathbf{P}})^2,\\
\label{eqn:3 order}
\mathbf{S}_{\rm NL}^{(3)}=&&-\frac{1}{{{e^2}n_0^2}}\Big( \nabla\cdot \mathbf{P}( \dot{\mathbf{P}}\nabla  \cdot \dot{\mathbf{P}}+ \dot{\mathbf{P}} \cdot \nabla \dot{\mathbf{P}}) + \dot{\mathbf{P}} \cdot \dot{\mathbf{P}}\nabla\nabla \cdot {\mathbf{P}}\Big) \nonumber\\
&&- \frac{1}{27}\frac{\beta^2}{e^2 n_0^2}\nabla (\nabla  \cdot {\mathbf{P}})^3.
\end{eqnarray}
\end{subequations}
In deriving these equations, contributions up to the third order have been considered. 
In particular, we used $n^{-1} \simeq n_0^{-1}\left( {1 - \frac{{{n_{{\mathop{\rm ind}}}}}}{{{n_0}}}} \right)$ and ${n^{2/3}} \simeq n_0^{2/3}\left[ {1 + \frac{2}{3}\frac{{{n_{{\mathop{\rm ind}} }}}}{{{n_0}}} - \frac{1}{9}{{\left( {\frac{{{n_{{\mathop{\rm ind}} }}}}{{{n_0}}}} \right)}^2}} \right]$.
As it can be noted, the nonlinear sources include both local and nonlocal terms. In particular, second-order terms include Coulomb, Lorentz, convective and quantum pressure contributions in that order in Eq.~\ref{eqn:2 order}, while third-order terms in Eq.~\ref{eqn:3 order} only include convective and pressure contributions. 

In the most general picture, for a single input field, THG is the process in which a signal of frequency $3\omega$ is generated  by the interaction of a field of frequency $\omega$ with a nonlinear material.
There are many ways, ideally infinite, to generate an output at $3\omega$.
In the most common process, three photons of energy $\hbar\omega$ combine to give a single photon of energy $3\hbar\omega$.
In this case, the THG is usually referred to as direct THG. 
Additionally the THG process can be seen as a two-steps process: first, a photon of energy $2\hbar\omega$ is generated by the interaction of two photons of energy $\hbar\omega$; then, this combines to another photons of energy $\hbar\omega$ in order to get a photon of energy $3\hbar\omega = 2\hbar\omega + \hbar\omega $.
The THG is then the combination of two second-order nonlinear processes, namely SHG followed by sum-frequency generation and the process is commonly defined as cascaded THG.
Being a combination of second-order processes, cascaded generation is forbidden in centrosymmetric bulk crystals, nevertheless, as we will show later, it is allowed in heavily doped semiconductors, similarly to metals  \cite{Celebrano:2015fy, DeLuca:2019, Celebrano:2019}, where second-order nonlinearities arise from FE dynamics. 

In order to describe the above mentioned process, let us assume time-harmonic dependence of the fields, i.e. ${\bf{F}}({\bf{r}},t) = \sum\limits_j {{{\bf{F}}_j}({\bf{r}}){e^{ - i{\omega _j}t}}}$, with ${\bf{F}}={\bf{E}}$, ${\bf{H}}$, or ${\bf{P}}$.
Eq.~\eqref{eqn:HM_P} with Eqs.~\eqref{eqn:NL_sources} and Maxwell's equations can be rewritten as a set of equations for each harmonic $\omega_j$ as follows:
\begin{subequations}
\label{eqn:HM_sys}
\begin{eqnarray}
&&\nabla\times\nabla\times \mathbf{E}_j-\varepsilon\frac{\omega_j^2}{c^2}\mathbf{E}_j-\omega_1^2\mu_0 (\mathbf{P}_j+\mathbf{P}_{\omega_j}^{\rm NL})=0,\\
&&{\beta ^2}\nabla (\nabla  \cdot {\mathbf{P}_j})+{(\omega_j^2+i\gamma\omega_j)\mathbf{P}_j= -\frac{{{n_0}{e^2}}}{m}\mathbf{E}_j \rule[-12pt]{0pt}{5pt}}_{\mbox{}} +\mathbf{S}_{\omega_j}
\end{eqnarray}
\end{subequations}
In writing these equations we have considered dielectric local contributions from the semiconductor, both linear, through the local permittivity $\varepsilon$, and nonlinear, through the nonlinear polarization $\mathbf{P}_{\omega_j}^{\rm NL}$.
Coupling between different harmonics occurs through the nonlinear contributions $\mathbf{P}_{\omega_j}^{\rm NL}$ and $\mathbf{S}_{\omega_j}$. 
For simplicity, we assume that the pump field is not affected by the nonlinear process (undepleted pump approximation), i.e. $\mathbf{P}_{\omega_1}^{\rm NL}=\mathbf{S}_{\omega_1}=0$, as we expect harmonic signals to be several orders of magnitude smaller than the pump fields. 
The system of Eqs.~\eqref{eqn:HM_sys} reduces then to three sets of one-way coupled equations, one for the fundamental ($j=1$), one for the second-harmonic frequency (j = 2) and one for the third-harmonic frequency ($j=3$).

The term that takes into account crystal lattice nonlinearities can be described using a bulk third-order susceptibility $\chi^{(3)}$ as:
\begin{eqnarray}
\label{eqn:NL_chi}
\mathbf{P}_{\omega_3}^{\rm NL}=\varepsilon_0\chi^{(3)}(\mathbf{E}_1\cdot\mathbf{E}_1)\mathbf{E}_1.    
\end{eqnarray}
In writing this equation we have assumed a centrosymmetric material, which also implies $\chi^{(2)}=0$.
The nonlinear source terms due to free charges are: 
\begin{eqnarray}
\label{eqn:SHG}
&&\mathbf{S}_{\omega_2}=-\frac{e}{m}\left({\mathbf{E}_1}\nabla  \cdot {\mathbf{P}_1}\right)- i\frac{e\mu_0}{m}\omega_1{\mathbf{P}_1} \times {\mathbf{H}_1}+\nonumber \\
&&\hspace{0.3cm}+\frac{\omega_1^2}{{e{n_0}}}( {\mathbf{P}_1}\nabla \cdot {\mathbf{P}_1} + {\mathbf{P}_1}\cdot \nabla {\mathbf{P}_1)}+ \nonumber\\
&&\hspace{0.3cm}-\frac{2}{3}\frac{\beta^2}{en_0}\left(\nabla  \cdot {\mathbf{P}_1}\nabla \nabla  \cdot {\mathbf{P}_1}\right),
\end{eqnarray}
for the SHG and $\mathbf{S}_{\omega_3}=\mathbf{S}_{\omega_3}^{(2)}+\mathbf{S}_{\omega_3}^{(3)}$ for the THG, with :
\begin{subequations}
\label{eqn:THG}
\begin{eqnarray}
\label{eqn:cTHG}
&&\mathbf{S}_{\omega_3}^{(2)}=-\frac{e}{m}\left({\mathbf{E}_2}\nabla  \cdot {\mathbf{P}_1}+{\mathbf{E}_1}\nabla  \cdot {\mathbf{P}_2}\right)+ \nonumber\\
&&\hspace{0.3cm}-i\frac{e\mu_0}{m}\left(\omega_2{\mathbf{P}_2} \times {\mathbf{H}_1}+\omega_1{\mathbf{P}_1} \times {\mathbf{H}_2}\right)+\nonumber \\
&&\hspace{0.3cm}+\frac{\omega_1\omega_2}{{e{n_0}}}( {{\mathbf{P}_2}\nabla \cdot {\mathbf{P}_1} + {\mathbf{P}_2}\cdot \nabla {\mathbf{P}_1} + {\mathbf{P}_1}\nabla \cdot {\mathbf{P}_2} + {\mathbf{P}_2}\cdot \nabla {\mathbf{P}_2}})\nonumber\\
&&\hspace{0.3cm}-\frac{2}{3}\frac{\beta^2}{en_0}\left(\nabla  \cdot {\mathbf{P}_2}\nabla \nabla  \cdot {\mathbf{P}_1}+\nabla  \cdot {\mathbf{P}_1}\nabla \nabla  \cdot {\mathbf{P}_2}\right),\\
\label{eqn:dTHG}
&&\mathbf{S}_{\omega_3}^{(3)}=-\frac{\omega _1^2}{e^2n_0^2}\Big[\nabla  \cdot \mathbf{P}_1( \mathbf{P}_1\nabla  \cdot \mathbf{P}_1+\mathbf{P}_1 \cdot \nabla \mathbf{P}_1 )+\nonumber\\
&&\hspace{0.3cm}+\mathbf{P}_1 \cdot \mathbf{P}_1\nabla \nabla  \cdot \mathbf{P}_1\Big] + \frac{1}{{27}}\frac{{{\beta ^2}}}{{{e^2}{n_0}^2}}\nabla (\nabla  \cdot {{\mathbf{P}}_1})^3
\end{eqnarray}
\end{subequations}
describing cascaded and direct THG due to FE dynamics, respectively. The hydrodynamic nonlinear sources in the general case of $\mathbf{S}_{\omega_1}\neq 0$ are reported in the appendix.
%

%%%%%%%%%%%%%%%%%%%%%%%%%%%%%%%%%%%%%%%%%%%%%%%%%%%%%%%%%%%%%%%%%%%%%%%%%%%%%%%%%%%%%%%%%%%%%%
%%%%%%%%%%%%%%%%%%%%%%%%%%%%%%%%%%%%%%%%%%%%%%%%%%%%%%%%%%%%%%%%%%%%%%%%%%%%%%%%%%%%%%%%%%%%%%

\section{\label{sec:results}Hydrodynamic nonlinearities in semiconductor slabs}
\subsection{Cascaded vs direct THG}
\begin{figure}[t]
	\centering
	\includegraphics[width=0.98\linewidth]{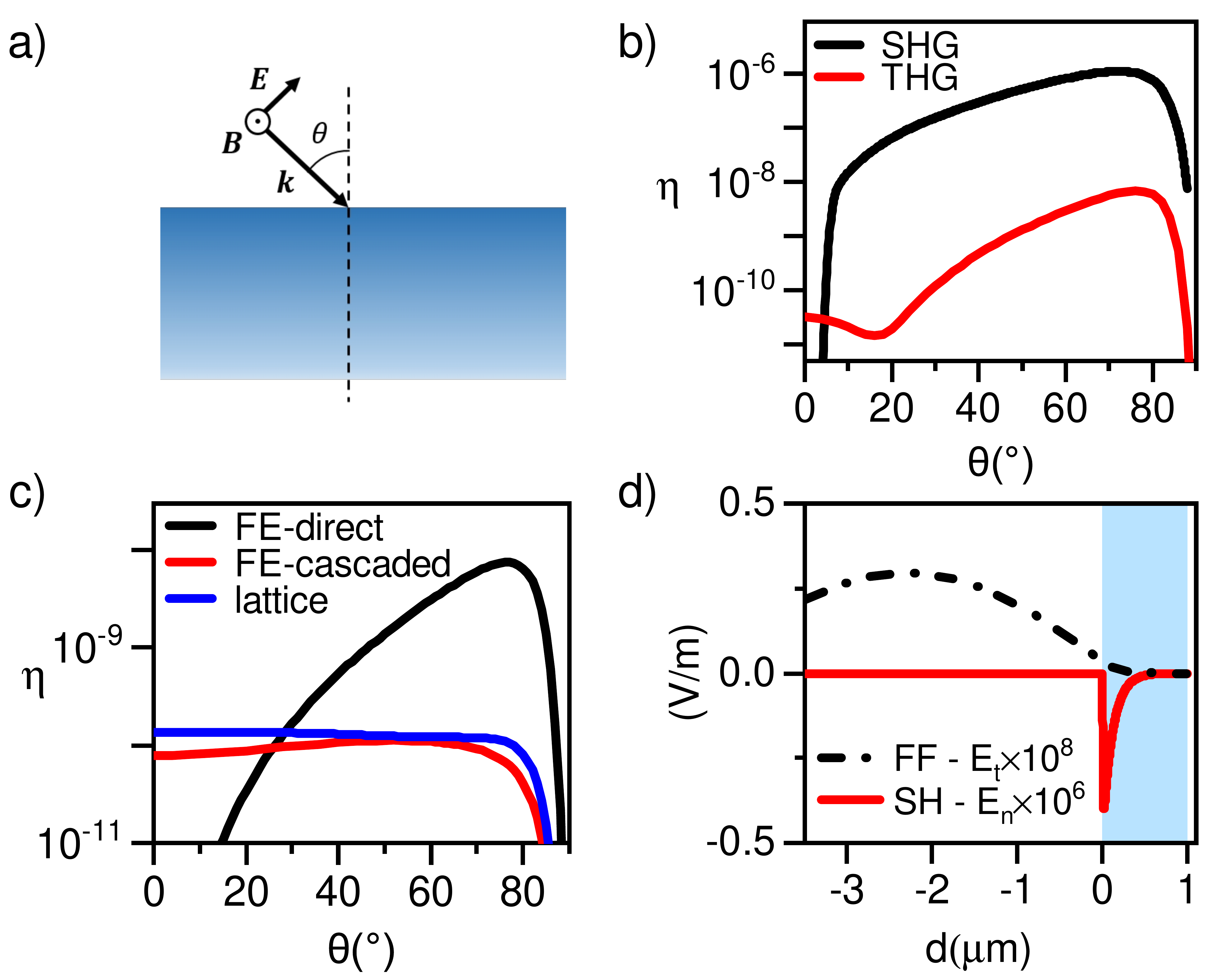}
	\caption{Hydrodynamic nonlinearities in a highly doped semiconductor slab: (a) a transverse magnetic (TM) plane wave impinging on the slab with an angle of incidence, $\theta$, is needed to induce a charge oscillation along the finite thickness of the slab; (b) SHG and THG efficiencies $\eta$ as a function of the incident angle for a InP slab (doping level $n_0 = 5.0 \times 10^{19}$~cm$^{-3}$, FF wavelength $\lambda_{\rm FF}=10~\um$); (c) THG efficiencies associated to FE direct, to FE cascaded and to lattice nonlinear contributions; (d) non-zero component of the fundamental (FF) and second-harmonic (SH) fields in the vicinity of the semiconductor surface as function of the distance $d$ from it. The simulated slab is $1~\um$ thick.}
	\label{fig:SH}
\end{figure}

An initial step toward the understanding of nonlinear conversion in heavily doped semiconductors is to consider a very simple geometry, namely a slab of a semiconductor.
Since a forcing electric field normal to the semiconductor surface is needed to induce a charge oscillation along the finite thickness of the slab, let us consider a transverse magnetic (TM) plane wave impinging on the slab with an angle of incidence, $\theta$, as depicted in Fig. \ref{fig:SH}a.

By solving Eqs.~(\ref{eqn:HM_sys}-\ref{eqn:THG}) we are able to calculate second- and third-harmonic generated signals.
We solved these equations numerically using the finite-element method (FEM) within a customized frequency-dependent implementation in COMSOL Multiphysics \cite{comsol}.
For our analysis, we consider Indium Phosphide (InP), a direct bandgap III-V semiconductor that has been recently studied as a low loss plasmonic material for the MIR region \cite{Panah:2016InP, Panah:2017s}, after being introduced among many alternative plasmonic materials few years ago \cite{Naik:2013am}.
InP is one of the most common materials in optoelectronic applications thanks to its easy integration and compatibility with the conventional microelectronic foundry and is particularly interesting for our study because of its very small effective electron mass, $m = 0.078 m_e$ \cite{Naik:2013am}, $m_e$ being the electron mass.
MIR properties of undoped InP can be described through a constant linear permittivity $\varepsilon_{\infty} = 9.55$ \cite{Naik:2013am} and a nonlinear bulk permittivity $\chi^{(3)} = 1.0 \times 10^{-18}$~m$^2$/V$^2$, a value that corresponds to the highest among most common semiconductors \cite{Boyd:2006uq}.
The pump peak intensity considered is $I_0 = 10$~W/\um$^2$ (1~GW/cm$^2$) in all cases and the THG efficiency has been obtained by normalizing the power of the generated signal to the input power at the fundamental frequency, $\eta=I_{\rm G}/I_0$, where $I_{\rm G}$ is the generated intensity.
The simulated slab is 1~\um~thick.

Figure~\ref{fig:SH}b shows SHG and THG efficiencies as a function of the incident angle, $\theta$, for a InP slab with doping level of $n_0 = 5.0 \times 10^{19}$~cm$^{-3}$ excited by a fundamental field (FF) oscillating at $\lambda_{\rm FF}=10~\um$.
As one would expect for a centrosymmetric material, the SHG signal at normal incidence is zero while it peaks at larger angles, $\theta\simeq72^\circ$, similarly to SHG from noble metals \cite{ODonnell:2005hc,RodriguezSune:2020uf}.
The THG efficiency, on the other hand, gives a non-zero contribution also at normal incidence, while increasing for large angles, peaking at $\theta\simeq77^\circ$.
As it can be inferred by Eq.~(\ref{eqn:dTHG}), direct FE contribution are zero at normal incidence, since $\nabla\cdot\mathbf{P}_1$ is identically zero for $\theta=0$.
Hence, it would be logical to think that the contribution shown in Fig.~\ref{fig:SH}b are due to the lattice nonlinearities, since also the SHG is zero.
In order to clarify this point, in Fig.~\ref{fig:SH}c we plot the THG efficiencies associated to each contributions.
Surprisingly the FE cascaded THG signal is not zero at normal incidence, moreover it is comparable to the lattice $\chi^{(3)}$ contribution.
How is it possible to obtain a cascaded effect from a signal that is apparently zero?
To understand this, we plot in Fig.~\ref{fig:SH}d the non-zero component of the fundamental and second-harmonic field in the vicinity of the semiconductor surface.
It can be observed indeed that the second-harmonic field has a non-zero longitudinal component that follows the profile of the square FF amplitude.
This component is induced by the Lorentz term in Eq.~(\ref{eqn:SHG}) inside the semiconductor.
However, because the vacuum does not support longitudinal modes, such component cannot be coupled to the far-field.
On the other hand, a second-harmonic longitudinal component in the semiconductor combines back with the magnetic field associated to the FF (through the Lorentz term in Eq.~(\ref{eqn:cTHG})), finally resulting in a transverse electric field component at the third harmonic.
A similar contribution is also given by the one of the convective terms.
Indeed, a THG cascaded contribution can be generated through an apparently zero SHG.    
\begin{figure*}[t]
	\centering
	\includegraphics[width=0.99\linewidth]{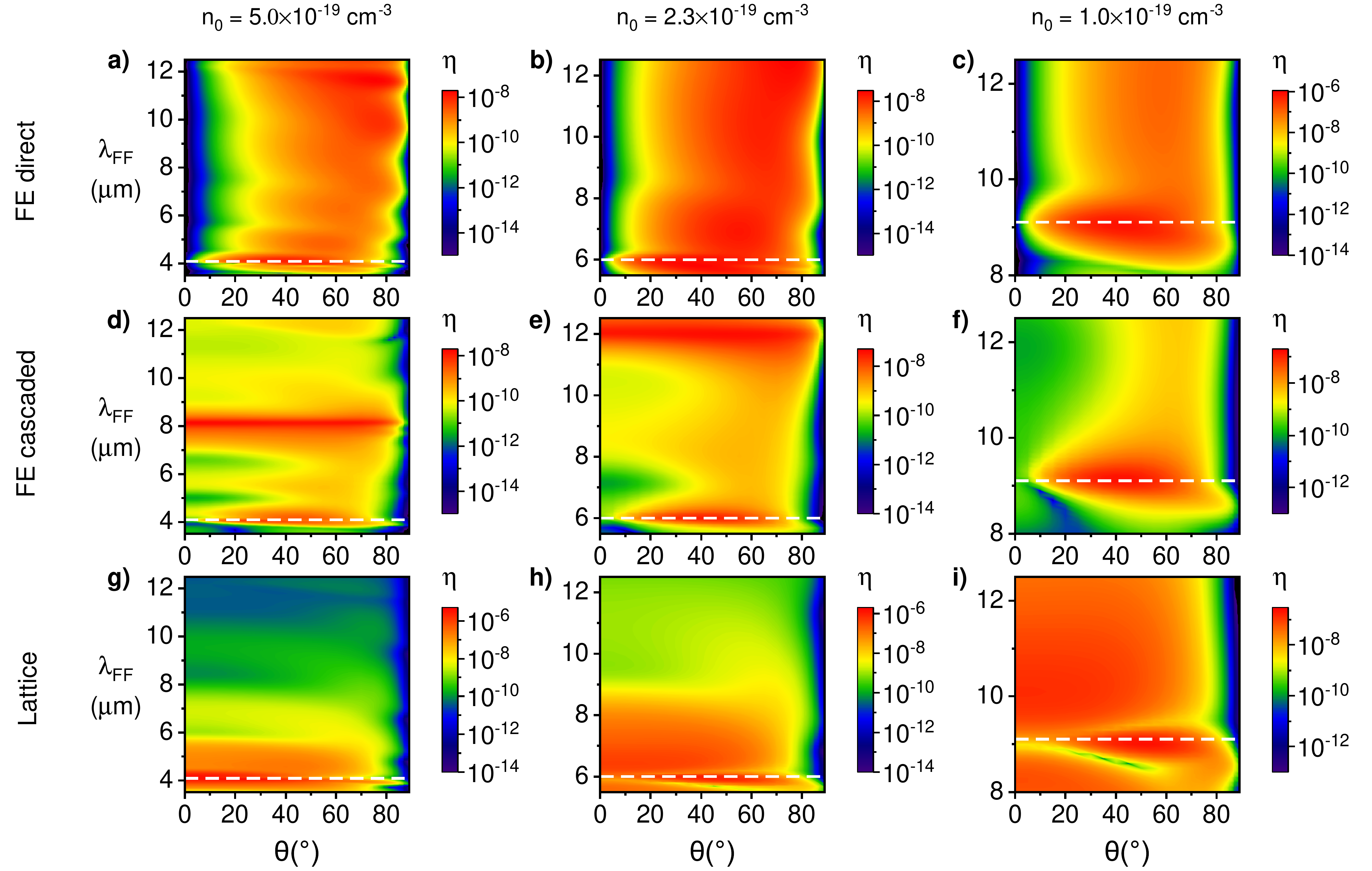}
	\caption{THG efficiency $\eta$ of a highly doped InP slab as a function of the incident angle $\theta$ and of the wavelength $\lambda_{\rm FF}$ of the FF, calculated taking into account distinct nonlinear sources: (a)-(c) are obtained considering only FE direct contributions ; (d)-(f) refer to the FE cascaded THG; (g)-(i) to the THG originating from background lattice third-order susceptibility $\chi^{(3)}$. In each column a different value (highlighted with a dotted white line) of the screened plasma wavelength, $\tilde{\lambda}_{\rm p}$ (i.e. a different doping level) is considered. In (a), (d) and (g) $\tilde{\lambda}_{\rm p} \simeq 4$~($n_0 = 5.0 \times 10^{19}$~cm$^{-3}$); in (b), (e) and (h) $\tilde{\lambda}_{\rm p} \simeq 6$~\um~($n_0 = 2.3 \times 10^{19}$~cm$^{-3}$); and in (c), (f) and (i)  $\tilde{\lambda}_{\rm p} \simeq 9$~\um~($n_0 = 1.0 \times 10^{19}$~cm$^{-3}$).}
	\label{fig:slab}
\end{figure*} 

Although this effect can be easily observed with numerical simulations, it is not that simple to differentiate between cascaded and direct contributions in an experimental set up.
Let us anticipate however that it is indeed possible to take advantage from this process and experimentally demonstrate this peculiar phenomenon by properly tuning the frequency with respect to the natural ENZ frequency of doped semiconductors, as we will show below.

\subsection{Hydrodynamic vs crystalline lattice contributions}
Having clarified the nature of cascaded effects, let us now focus on spectral properties of the THG process.
For a given material characterized by an effective mass $m$ and a fixed doping level $n_0$, the nature of its optical response depends on the frequency of the external electromagnetic field.
That is, the doped semiconductor has a metallic behavior for frequencies below the plasma frequency, while it behaves like a dielectric for higher frequencies.

In Fig.~\ref{fig:slab} each map plot reports the simulated reflected THG efficiency, $\eta$, of a slab of doped semiconductor as function of the FF incident angle, $\theta$, and wavelength, $\lambda_{\rm FF}$, for different conditions.
Specifically, each row refers to the THG efficiency obtained considering a distinct nonlinear source: $\mathbf{S}_{\omega_3}^{(3)}$ (FE direct), $\mathbf{S}_{\omega_3}^{(2)}$ (FE cascaded) and $\omega_1^2\mu_0\mathbf{P}_{\omega_3}^{\rm NL}$ (lattice) respectively.
Each column refers to a different level of doping, and, as a consequence, to a different screened plasma frequency, $\tilde{\omega}_{\rm p}\simeq {\omega}_{\rm p}/\sqrt{\varepsilon_\infty}$.

As discussed in the introduction, a higher $n_0$ is predicted to correspond to smaller hydrodynamic effects and thus to a smaller THG efficiency for a given ratio $\frac{\omega_{\rm FF}}{\tilde{\omega}_{\rm p}}$. 
For instance, if one considers the FE direct contribution, when $\lambda_{\rm FF} = \tilde{\lambda}_{\rm p}=\frac{2\pi c}{\tilde{\omega}_{\rm p}}$, $\eta$ goes from a peak of the order of $10^{-8}$ in Fig.~\ref{fig:slab}a to a maximum of $10^{-6}$ in Fig.~\ref{fig:slab}c, i.e. the efficiency is between 10 and 100 times bigger for a doping five times smaller.
Instead, for a given level of doping, an increase of the efficiency is expected when there are the most favourable energetic condition for the enhancement of the fields involved in the process.
Peaks of the efficiency are hence predicted for $\lambda_{\rm FF} = \tilde{\lambda}_{\rm p}$ and $\lambda_{\rm FF} = 3\tilde{\lambda}_{\rm p}$ in all cases, namely when the FF and the generated field are at ENZ condition, respectively.
These features can be clearly seen in all plots of Fig.~\ref{fig:slab} when the FF is at the ENZ condition.
However, when the generated signal is at the ENZ condition (only possible for the first column, for $\lambda_{\rm FF} \simeq 12$~\um), there is a sharp peak only for the FE direct THG (Fig.~\ref{fig:slab}a).
In the case of the cascaded process, there is no a clear maximum of $\eta$, probably because an enhancement of the generated field is not so effective for the overall efficiency if the other fields involved in the process are not enhanced as well.
The same happens for the lattice nonlinearities because of the limited interaction of the input field with the slab in the metallic regime of the material, since these contributions originate in the bulk.

A maximum of the efficiency for $\lambda_{\rm FF} = 2\tilde{\lambda}_{\rm p}$ is by contrast a unique feature expected only for the cascaded THG.
It corresponds to the enhancement of the field generated through SHG when it is at the ENZ condition.
The aforementioned peculiarity represents a crucial clue for the experimental demonstration of cascade effects in heavily doped semiconductors and can be clearly seen in Fig.~\ref{fig:slab}d and Fig.~\ref{fig:slab}e, for $\lambda_{\rm FF} \simeq 8$~\um~and $\lambda_{\rm FF} \simeq 12$~\um~, respectively.

A very important characteristic that emerges from Fig.~\ref{fig:slab}, is the order of magnitude of the FE THG compared to that arising from lattice nonlinearities. Indeed, although we used a very high value of $\chi^{(3)}$, third-order hydrodynamic contributions can be the most important in the portion of the MIR we considered.
This could be expected in the metallic regime of the material far from the screened plasma wavelength, where the interaction of the pump with the bulk crystal lattice is hindered.
However, for the smaller level of doping the efficiency of THG is higher for the hydrodynamic direct process than for the lattice even at ENZ.
This represents a remarkable prediction toward the application of heavily doped semiconductor for nonlinear plasmonics.

\begin{figure}[t]
	\centering
	\includegraphics[width=0.99\linewidth]{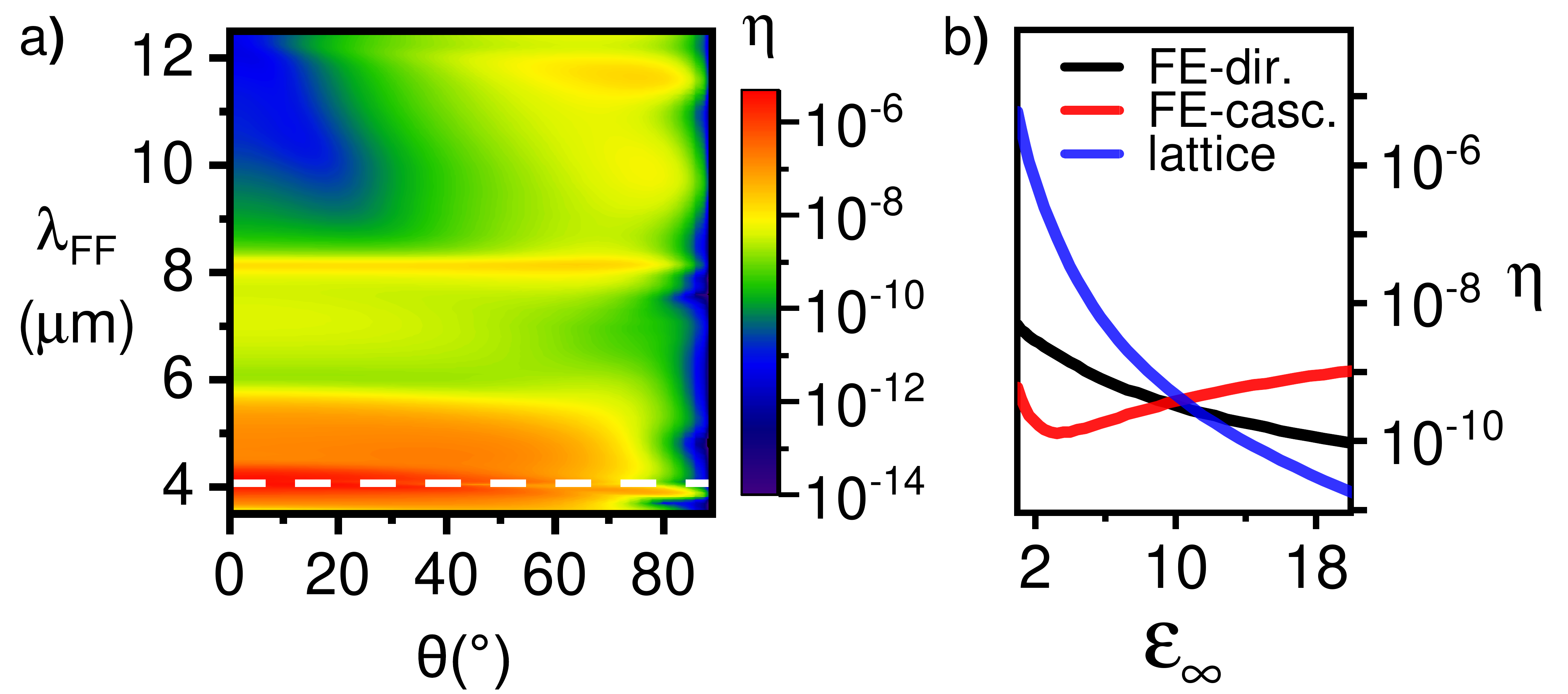}
	\caption{(a) THG efficiency $\eta$ of a highly doped InP slab as a function of the incident angle $\theta$ and of the wavelength $\lambda_{\rm FF}$ of the FF, calculated for $\tilde{\lambda}_{\rm p} \simeq 4$~($n_0 = 5.0 \times 10^{19}$~cm$^{-3}$) taking into account all the nonlinear sources; (b) THG efficiency as function of the background permittivity for the three distinct nonlinear contributions, with $n_0 = 5.0 \times 10^{19}$~cm$^{-3}$, $m = 0.1~m_e$, $\lambda_{\rm FF} = 1.8 ~ \tilde{\lambda}_{\rm p}$ and $\theta = 40^{\circ}$.}
	\label{fig:all_sources}
\end{figure}

\subsection{What to expect from experiments}
At this point of the discussion, it is interesting to wonder what kind of results one should expect from an experiment.
Let us now consider all the nonlinear sources at once to show what it is expected in a more realistic setup.
The results are reported in Fig.~\ref{fig:all_sources}a in the case of $\tilde{\lambda}_{\rm p} \approx 4$~\um~($n_0 = 5.0 \times 10^{19}$ cm$^{-3}$).
%We only report the efficiency, $\eta$, of the reflected THG, however, we encountered analogous features for the transmitted THG.
Unsurprisingly, in Fig.~\ref{fig:all_sources}a one can see all the main characteristics of the three nonlinear sources, which appear separately in the first column of Fig.~\ref{fig:slab}, combined in a single plot.
If our model is accurate enough  in describing nonlinearities in heavily doped semiconductors, an interesting way to demonstrate our predictions, taking into consideration Fig.~\ref{fig:all_sources}a, seems to look for a very peculiar trend of the THG efficiency at wavelengths close to $2\tilde{\lambda}_{\rm p}$ and to $3\tilde{\lambda}_{\rm p}$.
Comparing Fig.~\ref{fig:all_sources}a and Fig.~\ref{fig:slab}d, it is clear that the peak at $2\tilde{\lambda}_{\rm p}$ can only be caused by a cascaded generation.
A similar reasoning seems to make sense for the peak at $3\tilde{\lambda}_{\rm p}$ when considering the FE direct THG.

\subsection{The influence of the semiconductor background permittivity}
Hitherto, we considered a well-defined material, however, it is interesting to study how the hydrodynamic nonlinear sources are influenced by the material physical properties other the doping level $n_0$, even if the actual physical parameter cannot be varied in the experiments, as it is the case for the background permittivity, $\varepsilon_\infty$.
In Fig.~\ref{fig:all_sources}b we show the THG efficiency, $\eta$, as function of the $\varepsilon_\infty$ for the three THG contributions, with all other variables unchanged.
The trend of hydrodynamic THG efficiencies in Fig.~\ref{fig:all_sources}b might be intuitively understood thinking to the interaction of light with the slab from the point of view of Snell's law. 
Indeed, the refractive index increases with the square root of $\varepsilon_\infty$. This also defines the angle of refraction of a light inside a material (the higher is the refractive index, the smaller is the angle of refraction).
Therefore, for a given angle of incidence, a higher $\varepsilon_\infty$ causes the transmitted FF at the interface air/semiconductor to be closer to normal incidence when propagating inside the slab.
However, we know from the previous paragraphs that surface contributions are hindered at normal incidence, so, the FE direct THG efficiency, due only to surface contributions (see Eq.~(\ref{eqn:dTHG})) becomes smaller when increasing $\varepsilon_\infty$.
Conversely, the Lorentz term is maximum at normal incidence and it is then favoured for higher $\varepsilon_\infty$. As a consequence, taking into account Eq.~(\ref{eqn:cTHG}), the cascaded THG efficiency has a very peculiar trend caused by a balance between surface and Lorentz contributions.
Indeed, in Fig.~\ref{fig:all_sources}b, the FE cascaded contribution curve has the same decreasing trend as the FE direct contribution one for very small values of $\varepsilon_\infty$,
i.e. far from normal refraction, where surface contributions are dominant with respect to the Lorentz term.
It reaches a minimum where the contributions considered are of the same order and finally grows steadily with $\varepsilon_\infty$ when the Lorentz contribution becomes the most relevant (closer to normal refraction).
Instead, for background lattice nonlinearities, $\eta$ is higher for lower $\varepsilon_\infty$ simply because of the higher transparency of the material.
As a consequence, one should expect relatively higher hydrodynamic nonlinear contributions when considering a material with a lower background permittivity such as the conductive Indium Tin-Oxide (ITO)  \cite{Rodriguez:2020s}.

\begin{figure}[t]
	\centering
	\includegraphics[width=0.99\linewidth]{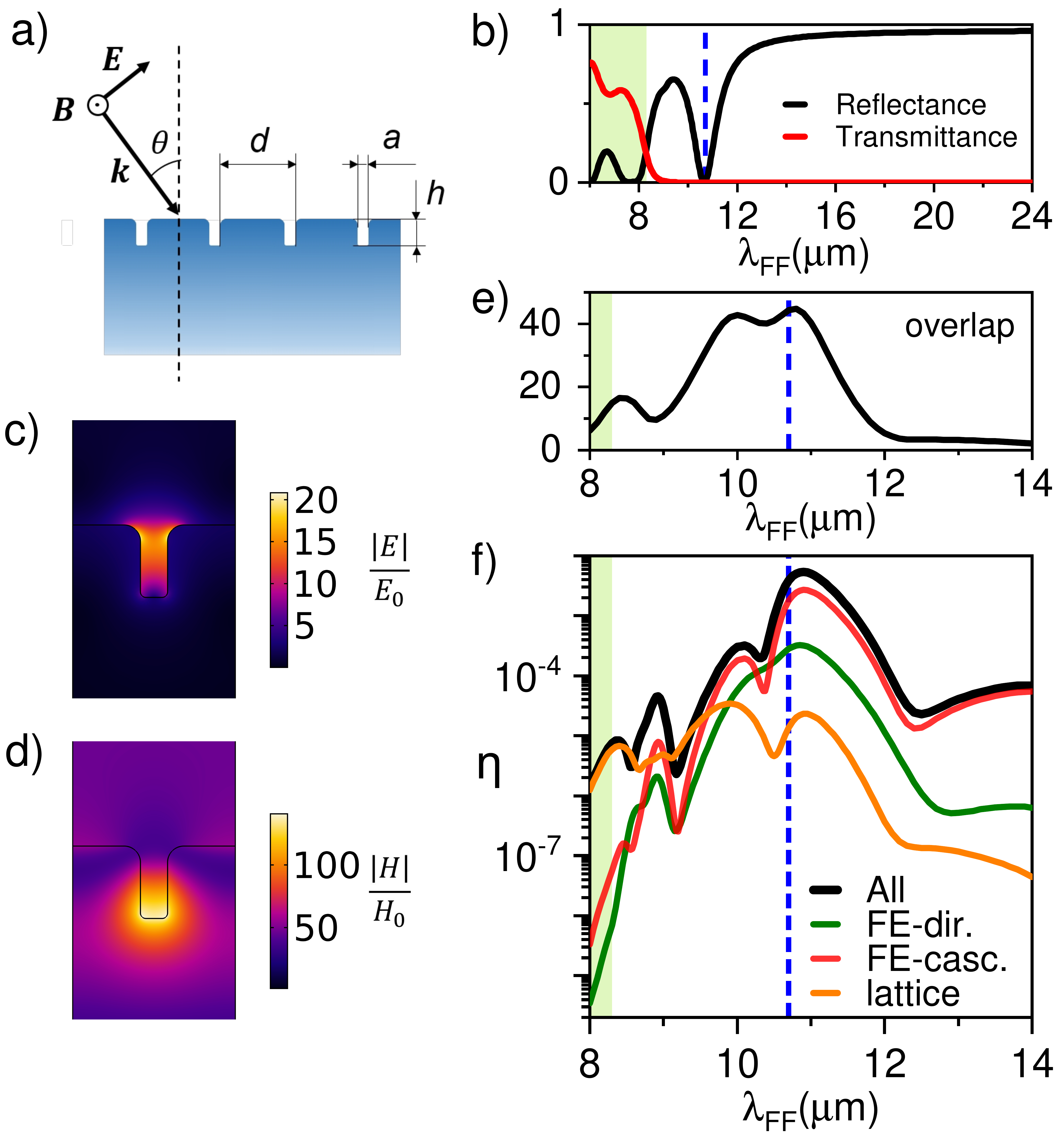}
	\caption{Optical properties of a InP grooves array for THG: (a) schematics of the geometry; (b) linear spectra of the structure at normal incidence when $n_0 = 1.2 \times 10^{19}$~cm$^{-3}$ ($\tilde{\lambda}_{\rm p} = 8.3~$\um), $d = 900$~nm, $a = 150$~nm and $h = 400$~nm, the thickness of the slab on which the grooves are carved is $H = 3$~ \um; (c-d) electric and magnetic field enhancements in correspondence of the resonance; (e) overlap integral as a function of the FF wavelength $\lambda_{\rm FF}$; (f) THG efficiency as function of $\lambda_{\rm FF}$ in the proximity of the resonance for the three distinct nonlinear contributions, FE direct, FE cascaded and  lattice and for their combination (All).}
	\label{fig:grooves}
\end{figure}

\section{A resonant pattern}

The analysis developed so far was useful to understand some fundamental properties of hydrodynamic nonlinearities. 
From the point of view of technology, however, it is perhaps more interesting to apply our model to a nanopatterned semiconductor slab with a geometrical resonance in the MIR.
To this aim, we take now into consideration an infinite array of parallel subwavelength grooves (a grating), as depicted in Fig.~\ref{fig:grooves}a.
This system supports plasmonic resonances and allows to couple virtually all incident energy into the active material \cite{LePerchec:2008ka,Pitelet:2019er,Dechaux:2016hg}, at normal incidence.
In Fig.~\ref{fig:grooves}b we report the simulated linear spectrum of the pattern we designed in order to have a resonance in the MIR for a TM polarized excitation.
Once the material properties have been fixed, the electromagnetic characteristics of the pattern can be designed as function of the array period $d$ and of the dimension of the grooves, the aperture $a$ and the height $h$ (see Fig.~\ref{fig:grooves}a).
The material considered is again InP and its doping level is $n_0 = 1.2 \times 10^{19}$~cm$^{-3}$ ($\tilde{\lambda}_{\rm p} = 8.3~$\um).
The pattern produces a resonance around $\lambda_{\rm res} = 10.6$~\um, where the reflectance is almost zero (see Fig.~\ref{fig:grooves}b).
In Fig.~\ref{fig:grooves}c and \ref{fig:grooves}d we report the electric and magnetic field enhancements in correspondence of the resonance. 
The fields are confined and enhanced into the grooves. This means that approximately all the pump energy impinging on the structure is coupled inside the material and localized in a much smaller volume.

In Fig.~\ref{fig:grooves}f, we report the reflected THG efficiency, $\eta$, at normal incidence, of the grooves array as function of the wavelength of the FF $\lambda_{\rm FF}$, considering the same processes analyzed for the slab: FE direct and FE cascaded, lattice and all nonlinear sources combined.
The pump peak intensity of the TM plane wave impinging on the structure  is the same as before in the case of the slab, $I_0 = 10$~W/\um$^2$ (1~ GW/cm$^2$).
A very important feature is the possibility to get FE direct THG even at normal incidence thanks to the plasmonic behaviour.
As it can be noted, all the curves, as expected, reach a peak approaching the resonance. Their trend is influenced by the overlap between the fundamental and generated fields, estimated calculating the overlap integral for the bulk contribution $\int \chi^{(3)} \mid \mathbf{E}_\omega \mid^2 \mathbf{E}_\omega\cdot\mathbf{E}^*_{3\omega} dV$, and reported in Fig.~\ref{fig:grooves}e (note that the exact overlap integral should take into account the specific expressions for all nonlinear sources).
The peak efficiency of FE contributions is several order of magnitude higher than in the case of the slab and it always surpasses that of intrinsic lattice nonlinearities. There, the maximum value was around $10^{-6}$ for the direct process and smaller for the cascaded, instead, here we get a value larger than $10^{-4}$ for the first process and bigger than $10^{-3}$ for the cascaded. Consequently, for the grooves array the efficiency is at least $10^2$ times larger for the FE direct THG and more than $10^4$ times higher for the cascaded THG, with the same pump intensity; while the THG due to the lattice increases just by a factor of 10.
Therefore, although we have considered a material with very high third-order nonlinear susceptibility, $\chi^{(3)}$, we predict that FE contributions will overcome by far conventional semiconductor nonlinearities in THG processes in the MIR.
  
%%%%%%%%%%%%%%%%%%%%%%%%%%%%%%%%%%%%%%%%%%%%%%%%%%%%%%%%%%%%%%%%%%%%%%%%%%%%%%%%%%%%%%%%%%%%%%
%%%%%%%%%%%%%%%%%%%%%%%%%%%%%%%%%%%%%%%%%%%%%%%%%%%%%%%%%%%%%%%%%%%%%%%%%%%%%%%%%%%%%%%%%%%%%%
\section{\label{sec:conclusions}Conclusions and perspectives}
We have used a hydrodynamic model to describe FE nonlinear optical dynamics in heavily doped semiconductors.
In particular, by deriving nonlinear hydrodynamic terms up to the third-order, we have investigated FE contributions to the THG process.
Contrarily to noble metals at optical frequencies, heavily doped semiconductor third-harmonic response is predicted to be strongly driven by hydrodynamic nonlinearities.
We have found in fact that such nonlinearities might generated THG signals up to two orders of magnitude larger than conventional semiconductor nonlinearities when combined to plasmonic field enhancement.
Moreover, we have showed that cascaded contributions, which are very often neglected, might be extremely important, even when one of the constituent signals appear to be negligible.
Although in most realistic experiments it would be impossible to differentiate cascaded effects from conventional nonlinear signals, we have proposed a simple method to experimentally verify our findings.
%These results not only have a fundamental relevance but can be crucial for future integrated MIR technologies.
Field-effect gated devices and optical generation of free carriers, both characteristics of semiconductors, could provide a route for tunable nonlinear optical systems on semiconductor-based photonic integrated circuit platforms.
We believe that these results could open new avenues for integrated nonlinear optics at MIR frequencies and beyond.

%\begin{acknowledgments}
%\end{acknowledgments}

\appendix
\begin{widetext}

\section{Hydrodynamic nonlinear sources}
In section~\ref{sec:model}, we derived the system of equations Eq.~(\ref{eqn:HM_sys}) and, assuming $\mathbf{S}_{\omega_1}= 0$, we reported in Eqs.~(\ref{eqn:SHG}-\ref{eqn:THG}) only the nonlinear sources $\mathbf{S}_{\omega_2}$ and $\mathbf{S}_{\omega_3}$. However, in the most general case $\mathbf{S}_{\omega_1}\neq 0$. It means that, for high input intensities, the nonlinear field produced may be converted back to the FF. Taking into account Eqs.(\ref{eqn:NL_sources}) this may happen through difference-frequency generation of the second order: 
\begin{eqnarray}
&&\mathbf{S}^{(2)}_{\omega_1= \omega_2-\omega_1}=-\frac{e}{m}\left({\mathbf{E}_1}\nabla  \cdot {\mathbf{P}^*_2} + {\mathbf{E}_2}\nabla  \cdot {\mathbf{P}^*_1} \right)- i\frac{e\mu_0}{m}\left(\omega_2{\mathbf{P}_2} \times {\mathbf{H}^*_1} -\omega_1{\mathbf{P}^*_1} \times {\mathbf{H}_2}\right)+\nonumber \nonumber\\
&&\hspace{0.3cm}+\frac{\omega_1^2}{{e{n_0}}}( {\mathbf{P}_2}\nabla \cdot {\mathbf{P}^*_1} + {\mathbf{P}_2}\cdot \nabla {\mathbf{P}^*_1}+{\mathbf{P}^*_1}\nabla \cdot {\mathbf{P}_2} + {\mathbf{P}^*_1}\cdot \nabla {\mathbf{P}_2})-\frac{2}{3}\frac{\beta^2}{en_0}\left(\nabla  \cdot {\mathbf{P}_2}\nabla \nabla  \cdot {\mathbf{P}^*_1} + \nabla  \cdot {\mathbf{P}^*_1}\nabla \nabla  \cdot {\mathbf{P}_2}\right),
\end{eqnarray}
through difference-frequency generation of the third order:
\begin{eqnarray}
&&\mathbf{S}^{(3)}_{\omega_1= \omega_3-\omega_1-\omega_1}= -\frac{{{\omega _1}^2}}{{{e^2}n_0^2}}\Big[ {(\nabla  \cdot {{\mathbf{P}}_3})\left( {{{\mathbf{P}}^*_1}\nabla  \cdot {{\mathbf{P}}^*_1} + {{\mathbf{P}}^*_1} \cdot \nabla {{\mathbf{P}}^*_1}} \right) + {{\mathbf{P}}^*_1} \cdot {{\mathbf{P}}^*_1}\nabla (\nabla  \cdot {{\mathbf{P}}_3})} \Big] + \nonumber\\
&&\hspace{0.3cm}+\frac{{{\omega _1}{\omega _3}}}{{{e^2}n_0^2}}\Big[{(\nabla  \cdot \mathbf{P}^*_1)\left( {{{\mathbf{P}}_3}\nabla  \cdot \mathbf{P}^*_1 + {{\mathbf{P}}_3} \cdot \nabla \mathbf{P}^*_1 + \mathbf{P}^*_1\nabla  \cdot {{\mathbf{P}}_3} + \mathbf{P}^*_1 \cdot \nabla {{\mathbf{P}}_3}} \right) + 2{{\mathbf{P}}_3} \cdot \mathbf{P}^*_1\nabla (\nabla  \cdot \mathbf{P}^*_1)} \Big] +\nonumber \\
&&\hspace{0.3cm} +\frac{1}{9}\frac{{{\beta ^2}}}{{{e^2}{n_0}^2}}\nabla \Big[{(\nabla  \cdot \mathbf{P}^*_1)^2}(\nabla  \cdot {{\mathbf{P}}_3})\Big],
\end{eqnarray}
eventually, third order effects may also occur as Kerr type nonlinearities:
\begin{eqnarray}
&&\mathbf{S}^{(3)}_{\omega_1= \omega_1-\omega_1+\omega_1}= -\frac{{{\omega _1}^2}}{{{e^2}n_0^2}}\Big[ (\nabla  \cdot {{\mathbf{P}}^*_1})\left({{\mathbf{P}}_1} \cdot \nabla {{\mathbf{P}}_1} \right) + {{\mathbf{P}}_1} \cdot {{\mathbf{P}}_1}\nabla (\nabla  \cdot {{\mathbf{P}}^*_1}) \Big] + \nonumber\\
&&\hspace{0.3cm}+\frac{\omega_1^2}{{{e^2}n_0^2}}\Big[{(\nabla  \cdot \mathbf{P}_1)\left({{\mathbf{P}}_1} \cdot \nabla \mathbf{P}^*_1 + \mathbf{P}^*_1\nabla  \cdot {{\mathbf{P}}_1} + \mathbf{P}^*_1 \cdot \nabla {{\mathbf{P}}_1} \right) + 2\mid\mathbf{P}_1\mid ^2\nabla (\nabla  \cdot \mathbf{P}_1)} \Big] + \nonumber\\
&&\hspace{0.3cm} +\frac{1}{9}\frac{{{\beta ^2}}}{{{e^2}{n_0}^2}}\nabla \Big[{\mid\nabla  \cdot \mathbf{P}_1\mid^2}(\nabla  \cdot {{\mathbf{P}}_1})\Big],
\end{eqnarray}
where all the variables and the constants have been introduced in sections~\ref{sec:intro}-\ref{sec:model}. As described in section~\ref{sec:model}, in deriving these equations we assumed time-harmonic dependence of the fields. 

\end{widetext}

%\nocite{*}


\end{document}